\documentclass[twocolumn,prl,superscriptaddress,showpacs]{revtex4}

\usepackage{epsfig}
\usepackage{bm}

\begin{document}
\date{\today}
\title{Magnetic friction in Ising spin systems}

\author{Dirk Kadau}
\affiliation{Dept.\ of Physics and CeNIDE, University of Duisburg-Essen, D-47048 Duisburg, Germany}
\affiliation{Institute for building materials, ETH Zurich, 8092 Zurich,
Switzerland}
\author{Alfred Hucht}
\author{Dietrich E. \surname{Wolf}}
\affiliation{Dept.\ of Physics and CeNIDE, University of Duisburg-Essen, D-47048 Duisburg, Germany}

\begin{abstract}
  A new contribution to friction is predicted to occur in systems with
  magnetic correlations: Tangential relative motion of two Ising spin
  systems pumps energy into the magnetic degrees of freedom. This
  leads to a friction force proportional to the area of contact. The
  velocity and temperature dependence of this force are investigated.
  Magnetic friction is strongest near the critical temperature, below
  which the spin systems order spontaneously. Antiferromagnetic
  coupling leads to stronger friction than ferromagnetic coupling with
  the same exchange constant. The basic dissipation mechanism is
  explained.  A surprising effect is observed in the ferromagnetically
  ordered phase: The relative motion can act like a heat pump cooling
  the spins in the vicinity of the friction surface.
\end{abstract}


\maketitle

As friction is an intriguingly complex phenomenon of enormous
practical importance, the progress in experimental techniques on the
micro- and nano-scale \cite{Baumberger,Frenken} as well as the
improved computational power for atomic simulations
\cite{Landmann,Sorensen,Holian} has led to a renaissance of this old
research field in recent years. Currently a large variety of
microscopic models compete with one another
\cite{Mueser,Persson,Baumberger}.  Major complications are wear,
plastic deformation at the contact, impurities, and lubricants. It is
unlikely that in the general case only a single dissipation mechanism
will be active. Defect motion, phononic and electronic excitations may
be involved in a very complex blend.
In order to reduce these complications and to focus on the elementary
dissipation processes, increasing attention has been paid to
non-contact friction: It can be measured as damping of an atomic force
microscope tip which oscillates in front of a surface without touching it
\cite{Baratoff,Moeller}. For this setup, too, phononic
\cite{Kantorovich1,Kantorovich2} as well as electronic dissipation
mechanisms \cite{Pendry,Volokitin} have been discussed. 
Recently, a Heisenberg model with magnetic dipole-dipole interactions
was studied at zero temperature as a model for magnetic force
microscopy. In this case  
the moving tip excites spin waves, which dissipate part of
the energy \cite{Fusco}.   

In this paper a different mechanism is considered, by which the spin
degrees of freedom of an Ising model contribute to friction.  We
imagine two magnetic materials with planar surfaces sliding on each
other.  Of course, if one of the materials is metallic, their relative
motion will induce eddy currents \cite{Hoffmann}. The corresponding
Joule heat is commonly associated with the term ``magnetic friction'',
although the energy is not dissipated into the spin degrees of
freedom, which can even be considered as frozen. By contrast, here we
are interested in the case that both materials are non-metallic (e.g.
magnetite Fe$_3\ $O$_4$). In order to highlight the role of the spin
degrees of freedom we do not take phononic and electronic excitations
into account explicitly, but regard them as a heat bath of
fixed temperature $T$ to which all spins are coupled.
Energy dissipation in Ising spin systems was studied previously
\cite{Acharyya,Ortin}, but there it was due to an oscillating magnetic
field rather than the tangential relative motion of two lattices. The
competition between the time scales for driving the system out of
equilibrium and for its relaxation gave rise to hysteretic, and hence
dissipative behavior. These time scales play also a role for
magnetic friction, as we will show.

Specifically, we present Monte-Carlo (MC) simulation results for a
two-dimensional Ising square lattice with periodic boundary
conditions.  Each of the $N$ lattice sites carries a classical spin
variable $S_i$ which can take the values $\pm 1$. The Hamiltonian is 
$H = -J\sum_{\langle i,j\rangle}S_i S_j$,
% \begin{equation}
%   \label{eq:ising}
%   H = -J\sum_{\langle i,j\rangle}S_i S_j,
% \end{equation}
where $\langle i,j\rangle$ denotes nearest neighbors, and $J$ is
chosen as energy unit.  Coupling to a
heat bath of constant temperature $T$ lets the spin configuration $C$
relax towards thermal equilibrium. The relaxation kine\-tics are 
determined by the transition rate $w(C\rightarrow C')$ to a new
configuration $C'$, in which one randomly chosen spin is flipped. We
consider fast relaxation with Metropolis rate \cite{LandauBinder}
\begin{equation}
w_{\rm M}(C\rightarrow C') = t_0^{-1} \min\left(1,e^{-\beta\Delta E}\right)
\end{equation}
and slow relaxation with Glauber rate \cite{LandauBinder}
\begin{equation}
w_{\rm G}(C\rightarrow C') = w_{\rm M}(C\rightarrow C') \left/\left(1+e^{-\beta
|\Delta E|}\right)\right.\ ,
\end{equation}
where $\beta=(k_{\rm B} T)^{-1}$. The energy difference
$\Delta E = E(C') - E(C)$ is received from ($\Delta E > 0$)
respectively transferred to ($\Delta E < 0$) the heat bath, when the
spin is flipped. $t_0 \approx 10^{-8} s$ 
\cite{Stearns} is the typical time for relaxation of a spin into the
direction of the local Weiss-field.

\begin{figure}[t]
\centerline{\epsfig{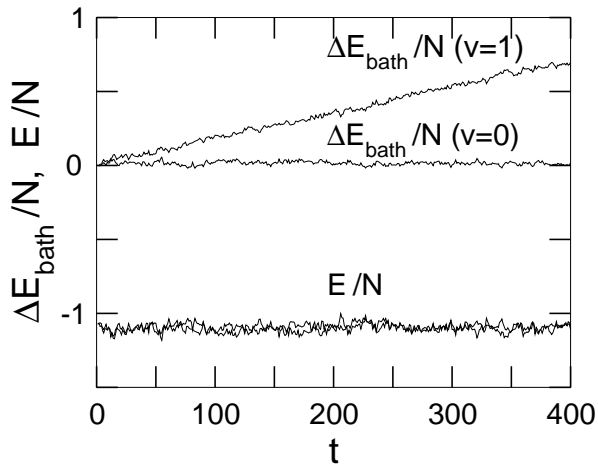}} 
%max7.7cm
\caption{Accumulated energy $\Delta E_{\rm bath}$ per spin which is
  transferred to the heat bath during a time interval $t$, without
  motion ($v=0$) and with motion ($v=1$) of the two half
  spaces. Simulation with Metropolis rates. The total energy per spin
  fluctuates around the exact value $E/N \approx -1.10608$
  \cite{McCoyWu} in both cases.} 
\label{fig:dissipation}
\end{figure}

The system is constantly driven out of equilibrium in the following
way: The lattice is cut parallel to an axis into an upper and a lower
half. The former is displaced by one lattice constant $a \approx 10^{-10} m$
in regular time intervals $a/v$, where $v$ is the sliding velocity (in
the following given in natural units $a/t_0$).  This means that $N/v$
random sequential spin updates (i.e. $1/v$ Monte Carlo steps) are
followed by a rigid translation of the upper half by one lattice
constant parallel to the cut.  $v=1$ corresponds to $10^{-2} m/s$.
(Note that due to the periodic boundary conditions there is a second
slip plane separating the upper half of the simulation cell from the
periodic image of the lower half.) The exchange interaction $J$ is the
same, no matter whether the interacting spins are on the same or on
different sides of the cut. This has the advantage that the relative
velocity $v$ and the temperature $T$ (in natural units $|J|/k_{\rm
  B}$) are the only parameters in the model. In the following we
evaluate the accumulated energy (divided by two, because of the two
equivalent slip planes) that has been exchanged with the heat bath
during the time interval $t$, $\Delta E_{\rm bath} (t)$, for different
sliding velocities $v$ and temperatures $T$. We first present our
results for ferromagnetic coupling, $J>0$.
In the end we also discuss, what is different for
antiferromagnetic coupling, $J<0$.

Is there any energy dissipation within this
simple model at all? To answer this question we first simulated a
system consisting of $80\times 80$ spins 
thermalized for 200 MC steps per spin at a
temperature $T=2.5$ above the
critical temperature $T_C=2/\ln(\sqrt 2 +1)$ \cite{McCoyWu} (initial
configuration). Figure \ref{fig:dissipation}
shows the energy exchange per spin with the heat bath for two cases: Without
relative motion ($v=0$) of the half spaces $\Delta E_{\rm bath}$ fluctuates
around $0$, i.e., no energy is dissipated. Switching on the relative
motion with a velocity $v=1$ leads to a
linear increase of $\Delta E_{\rm bath}(t)$. The total system energy
$E$ per spin stays constant at about the same value in both cases.
This means that the sliding system quickly develops a steady state,
where energy is transferred continuously to the heat
bath. The slope in Fig.\ref{fig:dissipation} is the constant
dissipation rate $P=\Delta E_{\rm bath}/\Delta t$. 
It is directly connected to the friction force $F$%:
by $P=F \ v$.
% \begin{equation}
%   \label{eq:Ploss}
%   P=F \ v
% \end{equation}
We conclude, that the Ising model gives rise to a truly magnetic
friction force: The relative motion pumps energy into the 
spin degrees of freedom, which in the steady state is then transferred
further into the heat bath. 

\begin{figure}
\centerline{\epsfig{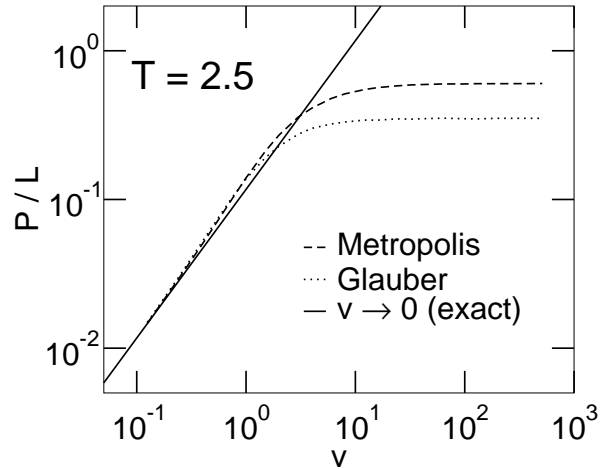}}
%max7.7cm
\caption{Energy dissipation rate $P$ per unit length  
  as function of the relative velocity $v$ of the two half spaces
  (averaged over 100 runs). Dashed line: Metropolis rates, dotted
  line: Glauber rates. Solid line: Exact solution for the limit $v
  \rightarrow 0$.}
\label{fig:dissipation_rate}  
\end{figure}

The magnetic friction force turns out to be proportional to the length
$L$ of the periodic cell along the direction of the cut through the
two dimensional lattice.  On the other hand, varying the system size
perpendicular to the slip plane does not change the above results, as
long as it remained larger than about 20 lattice constants. This shows
that whatever energy the relative motion pumps locally into the spin
degrees of freedom near the slip plane, gets transferred completely to
the heat bath before it can drive more distant parts of the system out
of equilibrium.

Figure \ref{fig:dissipation_rate} shows that the dissipation rate for
small velocities starts out linearly, with a slight upward curvature,
and saturates for large velocities. The saturation is expected, when
the velocity times the relaxation time $\tau$ becomes larger than the
correlation length $\xi$ \cite{def}, i.e. when $v > \xi/\tau$. Then the lower
half space is essentially confronted with uncorrelated configurations
of the upper half space, and a further increase of $v$ does not change
anything.
For Glauber dynamics the relaxation time is larger by a factor of
about 1.5 than for Metropolis dynamics. This explains the
difference between the curves in
Fig.~\ref{fig:dissipation_rate}: If one rescales time by this factor,
i.e. multiplies velocity and dissipation rate by 1.5, the
curve for Glauber dynamics is shifted such that it essentially
coincides with the one for Metropolis dynamics. 
For small velocities the linear $v$-dependence in
Fig.~\ref{fig:dissipation_rate} implies that the magnetic friction
force approaches a constant, $F_0$. For $T=2.5$ the velocity
independent part of the magnetic frictional shear stress has the value
$F_0/L = 0.114 \pm 0.004$. It is the same for Metropolis and Glauber
dynamics. 

\begin{figure}
\centerline{\epsfig{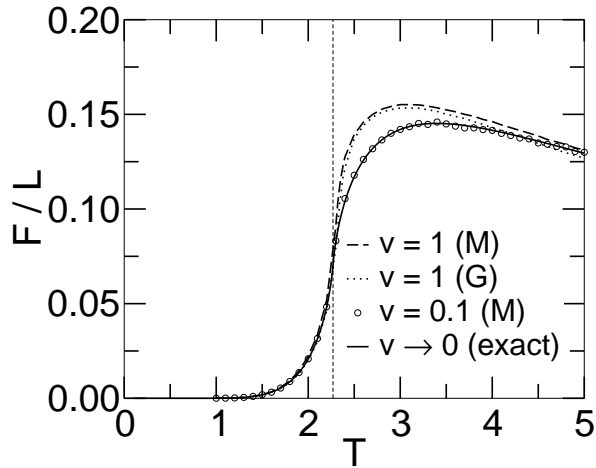}} 
%max7.7cm
\caption{ Temperature dependence of the friction force per unit
  length, $F/L$. Solid line: Exact quasistatic limit $v \rightarrow
  0$. Simulation results with Metropolis rates (circles) for $v=0.1$
  agree with the quasistatic limit. For $v=1$ the friction forces for
  Metropolis rates (dashed line) respectively Glauber rates (dotted line)  
  are larger corresponding to the upward curvature in
  Fig.~\ref{fig:dissipation_rate}. All data are averaged over 
  100 runs. The critical temperature is indicated by the dashed
  vertical line.} 
\label{fig:temperature}
\end{figure}

$F_0/L$ can be calculated analytically in the quasistatic limit,
$v\rightarrow 0$, where the spin system has time enough to relax back
into equilibrium after each displacement of the upper half. The energy
of the spin configuration immediately after a displacement minus the
equilibrium energy must be transferred to the heat bath during the
time interval $a/v$. The rigid shift of all spins in the equilibrated
upper half by one lattice constant places former next nearest
neighbors in nearest neighbor positions on opposite sides of the slip
plane. Thus the dissipated energy per unit length $a$ (i.e. the
friction force) can simply be expressed as $JL$ times the nearest
neighbor spin correlation function minus the next nearest neighbor spin
correlation function. Both are known analytically, see e.g. Eq.(4.5)
and (4.9) of \cite{McCoyWu}. At $T=2.5$ this gives the value $F_0/L
\approx 0.117$ in good agreement with the numerical result. For
general temperature one obtains the solid curve in Fig.
\ref{fig:temperature}.

According to the picture of Bowden and Tabor \cite{BowdenTabor} also
Coulomb friction is independent of $v$ and proportional to the real
contact area, which due to surface roughness is smaller than the
sliding surface macroscopically appears to be, and grows proportional
to the normal load. Therefore, the velocity independent part of the
magnetic friction force behaves like Coulomb friction.  How does it
compare to typical values for solid friction?  The above results show
that 
the  magnetic shear stress % change acc. to ref B is that enough??
 $\sigma_{\rm t}=F_0/L$ is of the order of 0.1 for the two
dimensional Ising model. The unit is $J/a^2$, the exchange constant
divided by the lattice constant squared. If we regard the two
dimensional Ising model as a slice of thickness $a$ of a three
dimensional system, then we may assume that the magnetic shear stress
for a three dimensional Ising model is of the order of $\sigma_{\rm t,
  3d} \approx 0.1 J/a^3$.  Inserting typical values ($J \approx 0.6
\cdot 10^{-20}$ Joule, $a \approx 3\cdot 10^{-10}$m) one gets the
estimate $\sigma_{\rm t, 3d} \approx 20\ {\rm MPa}$.  This is a
surprisingly large value.  Ordinary solid friction shear stresses are
given by $\sigma_{\rm t, Coulomb} = \mu \sigma_{\rm c}$ according to
the Bowden-Tabor-theory, where a typical value for the friction
coefficient is $\mu = 0.2$, and the yield stress $\sigma_{\rm c}$ at
high temperatures is a few hundred to thousand MPa. We conclude that
magnetic friction is probably not too weak compared to ordinary solid
friction to be observable.

There is one caveat, however: The exchange interaction is extremely
short range, but in the simulation results presented here no reduced
value was inserted for 
the interaction of spins on opposite sides of the slip plane. The
above estimate should therefore only be applied, if the surfaces are
in close contact. As expected, simulations with a reduced magnetic
exchange interaction across the slip plane lead to a smaller friction force.

Magnetic friction has characteristic features near the critical
temperature, which should be useful to separate this contribution to
solid friction from other ones. It is
nearly zero at low temperatures, where the ferromagnetic ordering
implies almost perfect translational invariance along the surface. As
thermal fluctuations destroy the translational invariance, magnetic
friction raises sharply to a maximum slightly above the critical
temperature (Fig.~\ref{fig:temperature}). In the paramagnetic region 
the exact quasistatic limit shows that the friction force has the same
$1/T$ asymptotics as % like %change acc. ref B
$JL$ times the nearest neighbor correlation
function, because the next nearest neighbor correlation ($\propto
1/T^2$) becomes negligible.

\begin{figure}
\centerline{\epsfig{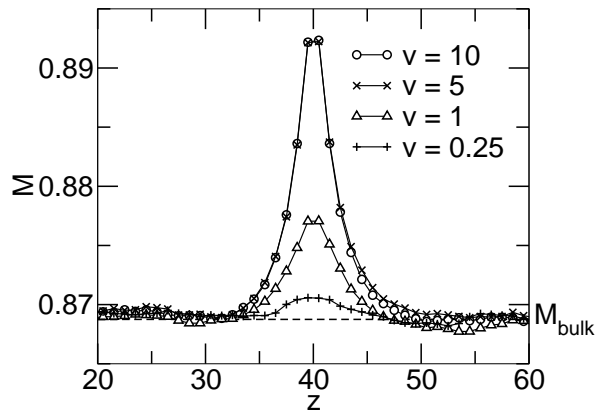}} %max7.7cm
\caption{Magnetization profile along the $z$-axis perpendicular to the
  slip plane (at $z=40$ in units of the lattice constant $a$) at
  $T=2.1<T_C$ for Glauber rates at different velocities.
%$v=0.25$ ($+$), $v=1$ ($\triangle$), $v=5$ ($\times$), and $v=10$ ($\bigcirc$).
The local magnetization near the slip plane is
enhanced. This effect becomes stronger for increasing velocity and
saturates for the same reason as in Fig.~\ref{fig:dissipation_rate}. 
When using Metropolis rates this effect is less pronounced (not shown).}
\label{fig:magnetization_profile}
\end{figure}

What is the basic mechanism leading to magnetic friction in the Ising model?
Obviously, shearing reduces the correlation length locally
by disturbing the equilibrium correlations between spins
on opposite sides of the slip plane. 
Above $T_c$ this corresponds to an effective temperature
increase, which explains the energy flow into the cooler heat
bath. Since more neighbor pairs with antiparallel spin are present, 
the energy density is locally increased in the steady state, compared to its 
value in thermal equilibrium.
As the correlation length vanishes for $T \rightarrow \infty$, this
picture explains why magnetic friction vanishes in this limit.

Below $T_c$ the correlation length can be associated with the diameter
of thermally activated minority clusters of spins pointing into the
direction opposite to the spontaneous magnetization.  The relative
motion distorts minority clusters, which extend across the slip plane,
and possibly cuts them into two pieces.  Again this reduces the
effective correlation length. In thermal equilibrium a smaller
correlation length indicates a better ordered magnetic state. Indeed
we find an increased magnetization locally at the slip plane
(Fig.~\ref{fig:magnetization_profile}). This effect is less pronounced 
for the Metropolis algorithm, where the
spin configurations relax more quickly into
thermal equilibrium. 

The local spin temperature in the
vicinity of the slip plane drops due to the influence of shearing. The
driven system acts like a ``heat pump'' cooling the spin degrees of
freedom below the temperature of the heat bath. The shearing creates additional
domain walls by deforming or fragmenting minority clusters. 
The system continuously tries to reduce these excess domain walls, 
thereby transferring domain wall energy to the heat
bath. This is the dissipation mechanism. 

Why does this ``heat pump'' work better for higher velocities, as
shown by Fig.~\ref{fig:magnetization_profile}? Let us
discuss first the case of sufficiently high velocities, where the
magnetization near the slip plane saturates at a maximal value. 
Then correlations between the two half spaces can be neglected.
Instead, the spins in
the lower half see an effective surface field corresponding to the
average surface magnetization of the upper half. Hence minority spins
near the slip plane flip more easily into the majority direction than
in the bulk. For smaller velocities, however,  minority clusters can be
stabilized more and more because of correlations across
the slip plane. Hence the surface magnetization decreases. 

Analogous investigations for antiferromagnetic coupling ($J<0$) were
done, too. The dissipation rate turns out to be much higher than in
the ferromagnetic case (with the same $|J|$).  The friction maximum is
more than three times larger for the Ising antiferromagnet than for
the ferromagnet.  The reason is that the local antiferromagnetic order
across the slip plane is destroyed, whenever the upper lattice is
displaced by one lattice constant. This is a stronger perturbation
than in the ferromagnetic case, where only the correlations of thermal
disorder could be destroyed by the relative motion. In particular
magnetic friction does not vanish for $T\rightarrow 0$ in the
antiferromagnetic case.

We acknowledge funding by the DFG through SFB 616 (``Energy
dissipation at surfaces'') and support by Federal Mogul Technology GmbH.


\begin{thebibliography}{99}
\bibitem{def} The equilibrium bulk correlation length $\xi$ is defined
  by the exponential cutoff of the spin-spin correlation function, 
  $\langle S_i S_j \rangle - \langle S \rangle^2 \sim
  |i-j|^{-1/2}\exp(-|i-j|a/\xi)$ for large $|i-j|$ and $T\neq T_c$
  (Ornstein-Zernike-form, \cite{LandauBinder} eq.\ 2.22).
\end{thebibliography}
\end{document}